\documentclass[prl,twocolumn,superscriptaddress,showpacs,unsortedaddress]{revtex4}
\usepackage{graphicx}
\usepackage{dcolumn}
\usepackage{amsmath}

\begin{document}

\title{Triple-$\vec{q}$ octupolar ordering in $\mathrm{NpO_2}$}

\author{J. A. Paix{\~a}o}

\affiliation{Departamento de F{\'\i}sica, Universidade de Coimbra,
  P-3004 516 Coimbra, Portugal}

\author{C. Detlefs}

\affiliation{European Synchrotron Radiation Facility, Bo\^{\i}te
  Postale 220, F--38043 Grenoble Cedex, France}

\author{M. J. Longfield}

\affiliation{European Commission, Joint Research Centre, Institute for
  Transuranium Elements, Postfach 2340, D--76125 Karlsruhe, Germany}

\author{R. Caciuffo}

\affiliation{Istituto Nazionale per la Fisica della Materia,
  Dipartimento di Fisica ed Ingegneria dei Materiali, Universit\`a di
  Ancona, Via Brecce Bianche, I--60131 Ancona, Italy}

\author{P. Santini}

\affiliation{Oxford Physics, Clarendon Laboratory, Oxford OXI 3PU,
  United Kingdom}

\author{N. Bernhoeft}

\affiliation{D\'epartment de Recherche Fondamentale sur la Mati\`ere
  Condens\'ee, Commissariat \`a l'\'Energie Atomique-Grenoble,
  F--38054 Grenoble, France}

\author{J. Rebizant}
\author{G. H. Lander}

\affiliation{European Commission, Joint Research Centre, Institute for
  Transuranium Elements, Postfach 2340, D--76125 Karlsruhe, Germany}

\date{Version \today}


\begin{abstract}
  We report the results of resonant X-ray scattering experiments
  performed at the $\mathrm{Np}$ $M_{4,5}$ edges in $\mathrm{NpO_2}$.
  Below $\mathrm{T_0 = 25~K}$, the development of long-range order of
  $\mathrm{Np}$ electric quadrupoles is revealed by the growth of
  superlattice Bragg peaks.  The electronic transition is not
  accompanied by any measurable crystallographic distortion, either
  internal or external, so the symmetry of the system remains cubic.
  The polarization and azimuthal dependence of the intensity of the
  resonant peaks is well reproduced assuming anisotropic tensor
  susceptibility (ATS) scattering from a triple-$\vec{q}$ longitudinal
  antiferroquadrupolar structure.  Electric quadrupole order in
  $\mathrm{NpO_2}$ could be driven by the ordering at $\mathrm{T_0}$
  of magnetic octupoles of $\Gamma_{5}$ symmetry, splitting the
  $\mathrm{Np}$ ground state quartet and leading to a singlet ground
  state with zero dipole magnetic moment.
\end{abstract}

\maketitle

For half a century the low temperature properties of $\mathrm{NpO_2}$
have mystified theorists and experimentalists alike.  Upon cooling
from room temperature a \emph{single} phase transition is observed at
$\mathrm{T_0 \approx 25.5~K}$ \cite{Osborne53}.  Similarities with
$\mathrm{UO_2}$ \cite{Caciuffo99} suggested a magnetic nature of the
phase transition, but no magnetic order was found by neutron
diffraction \cite{Caciuffo87} nor by M{\"o}{\ss}bauer spectroscopy
\cite{Friedt85}, which established an upper limit of $\approx
0.01~\mu_{\mathrm{B}}$ for the \emph{ordered} magnetic moment
$\vec{\mu}_{\mathrm{ord}}$. However, $\mathrm{Np}^{4+}$ ions in
$\mathrm{NpO_2}$ are Kramers ions ($5f^{3}$, $^4I_{9/2}$).  In the
absence of interactions breaking time-reversal symmetry the ground
state has to carry a magnetic moment, $\vec{\mu}$, whatever the
crystalline environment. A fluctuating magnetic moment of finite size
would be revealed by a Curie-like divergence of the susceptibility at
low temperatures, whereas the experiments reveal a flat susceptibility
between $\mathrm{15}$ and $\mathrm{5~K}$ \cite{Erdoes80}. Moreover, no
evidence for a crystallographic distortion, neither external nor
internal, has been found by synchrotron experiments \cite{Mannix99}.

Another element of the puzzle was recently provided by muon spin
relaxation experiments, showing the abrupt appearance of a precession
signal below $\mathrm{T_0}$ \cite{Kopmann98}. This implies that the
order parameter (OP) sets up a magnetic field at the muon stopping
site and provides definitive evidence that the OP breaks invariance
under time reversal. By assuming antiferromagnetic (AF) order of the
same kind of that established for $\mathrm{UO_2}$, i.e.{\ }a type-I,
triple-$\vec{q}$ structure, the authors deduced an ordered moment
$\mu_{\mathrm{ord}} \approx 0.1~\mu_{\mathrm{B}}$, a value much larger
than the upper limit compatible with M{\"o}{\ss}bauer spectroscopy.
In parallel with this finding, direct evidence for long-range order in
$\mathrm{NpO_2}$ was obtained through the observation below
$\mathrm{T_0}$ of superlattice reflections in resonant X-ray
scattering (RXS) experiments at the $\mathrm{Np}$ M$_{4,5}$ absorption
edges \cite{Mannix99}. The superstructure Bragg peaks occur at
$\vec{Q}=\vec{G} + \langle 0~0~1 \rangle$ positions, where $\vec{G}$
is a reciprocal lattice vector. This is the same periodicity found for
the AF phase in $\mathrm{UO_2}$, and the observations were taken as
evidence for the occurrence of longitudinal triple-$\vec{q}$ AF order.
Santini and Amoretti \cite{Santini00} pointed out the possibility of
explaining the whole body of experimental evidence assuming
magnetic-octupole order instead of magnetic-dipole order.  Octupolar
order would lift the degeneracy of the $\Gamma_{8}$ $\mathrm{Np}$
ground state and generate an interstitial magnetic field, in agreement
with neutron spectroscopy \cite{Amoretti92} and muon spin resonance
results.  However, octupolar order can be directly observed in RXS
only through E2-resonances, whilst the resonances observed in
\cite{Mannix99} are at the E1 absorption edge.  Indeed, all previous
experimental data taken at the actinide M$_{4,5}$ edges indicates a
very strong dominance of E1-processes ($3d_{3/2,5/2} \leftrightarrow
5f$). To our knowledge there is no evidence for E2 contributions
($3d_{3/2,5/2} \leftrightarrow 6g$), and indeed matrix elements
involving E2 promotion to $6g$ states are expected to be too small to
allow for any observable signal.

To clarify the above confusion, we have undertaken a new RXS
experiment at the magnetic scattering beamline, ID20, of
the ESRF\@.  We performed polarization analysis of the diffracted
radiation and measured the dependence of the intensity from the
azimuthal angle, $\psi$ (the angle describing the rotation of the
crystal about the scattering vector). These analyzes were not possible
in the experiment reported earlier in \cite{Mannix99}, preventing an
unambiguous determination of the origin of the resonance. Indeed, the
results we present in this Letter show that the superlattice peaks in
$\mathrm{NpO_2}$ are not magnetic, but arise from the asphericity of
the $\mathrm{Np}$ $5f$ electron density leading to an anomalous tensor
component in the atomic scattering factor.  In other words, the
superlattice peaks signal the occurrence of electric-quadrupole long
range order below $\mathrm{T_0}$. 
This conclusion is incompatible with the particular octupolar model
given in Ref.~\cite{Santini00}, which predicts an undistorted charge
density for the $5f$ ground state, with vanishing quadrupole moment.

The experiment was performed on a single crystal of $0.7\times
0.7\times 0.2 \mathrm{mm}^{3}$ in volume, with a flat $(0~0~1)$
surface. A closed-cycle refrigerator equipped with an
azimuthal-rotation stage provided a base temperature of
$\mathrm{12~K}$. An $\mathrm{Au} (1~1~1)$ single crystal was used to
analyze whether the polarization of the scattered beam was parallel
($\pi$) or perpendicular ($\sigma$) to the scattering plane.  The
incident beam had $\sigma$ polarization and the scattering plane was
vertical.

\begin{figure}
  \centerline{%
    \includegraphics[angle=90,width=0.75\columnwidth]{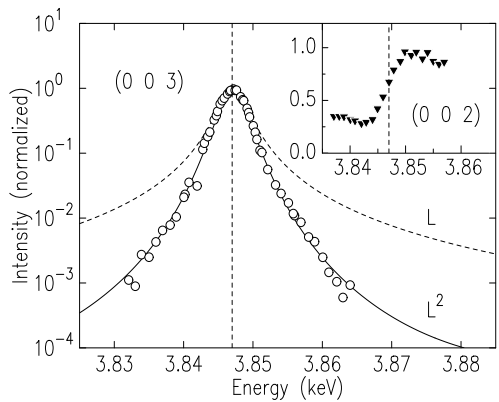}
    }
  \caption[Energy scans through the $\mathrm{Np}$
  M$_4$ absorption edge.]{\label{fig.escan}Energy scans through the
    $\mathrm{Np}$ M$_4$ absorption edge. The $\sigma\rightarrow\pi$
    (main panel) and $\sigma\rightarrow\sigma$ (not shown) resonances
    of the $(0~0~3)$ superstructure peak have a Lorentzian-squared
    (solid line) rather than a Lorentzian (dashed line) line shape.
    The maximum at the resonance lies at the absorption edge as
    measured at the $(0~0~2)$ Bragg reflection (insert).}
\end{figure}

Fig.~\ref{fig.escan} shows the $\sigma\rightarrow\pi$ intensity of the
$\vec{Q}=(0~0~3)$ superlattice reflection as a function of the photon
energy, $\mathrm{E}$, around the $\mathrm{Np}$ $M_{4}$ absorption
edge. The data can be fit to a Lorentzian squared line-shape, centered
near the E1 threshold. The presence of a resonance in the
$\sigma\rightarrow\sigma$ channel (not shown) excludes that the
scattering arises from dipole-magnetic order, as this would give only
$\sigma\rightarrow\pi$ resonant scattering. Studies measuring magnetic
order at actinide M edges have shown Lorentzian energy profiles,
whereas we clearly observe a Lorentzian-squared form, as it is
expected when the intermediate state splitting is much smaller than
the core hole life time \cite{Zimmermann01}.

\begin{figure}
  \centerline{%
    \includegraphics[height=0.52\textheight]{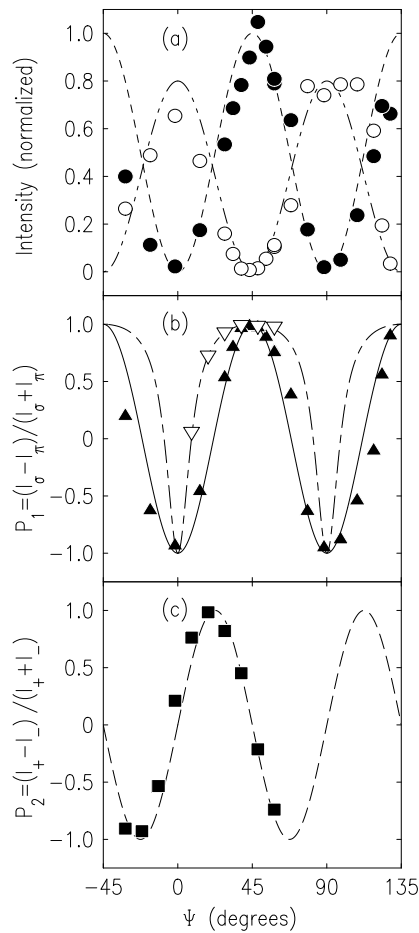}
    }
  \caption[Dependence of the scattering on the azimuthal angle,
  $\psi$.]{\label{fig.intensities}Dependence of the scattering on the
    azimuthal angle, $\psi$: (a) Intensity at the $(0~0~3)$
    quadrupolar reflection in the $\sigma$- (filled circles) and
    $\pi$- (open circles) channels with incident polarization
    $\sigma$. Different scaling factors have been applied to correct
    for differences in the efficiency of the PA in the $\sigma$ and
    $\pi$ settings. (b) Stokes parameter $P_1$ for the $(0~0~1)$ (open
    triangles) and $(0~0~3)$ (filled triangles) quadrupolar
    reflections.  (c) Stokes parameter $P_2$ for the $(0~0~3)$ peak.
    The lines represent model calculations based on
    eqs.~\ref{eq.sigma},~\ref{eq.pi}, and~\ref{eq.plusminus} }
\end{figure}

The results of $\psi$ (azimuthal) scans about the $(0~0~3)$ peak,
taken at $\mathrm{T = 12~K}$ with $\mathrm{E = 3.846~keV}$, are shown
in Fig.~\ref{fig.intensities}(a) for both the
$\sigma\rightarrow\sigma$ and the $\sigma\rightarrow\pi$ channel.
Fig.~\ref{fig.intensities}(b) and~\ref{fig.intensities}(c) show,
respectively, the $\psi$ dependence of the Stokes parameter
$P_1=(I_{\sigma\rightarrow\sigma} -
I_{\sigma\rightarrow\pi})/(I_{\sigma\rightarrow\sigma} +
I_{\sigma\rightarrow\pi})$ and $P_2 =(I_{+}-I_{-})/(I_{+}+I_{-})$
\cite{Lipps54}, for the $(0~0~1)$ and $(0~0~3)$ reflections. $P_2$ can
be considered as a measure of the phase relation between the two
polarization channels and is defined by the intensities measured with
the analyzer oriented at $\pm 45^\circ$ with respect to the scattering
plane,
\begin{equation}
     I_{\pm} \propto \frac{1}{\sqrt{2}}
     \left|F_{\sigma\rightarrow\sigma} \pm F_{\sigma\rightarrow\pi}\right|^2
     \label{eq.one}
\end{equation}
where $F_{\sigma\rightarrow\sigma}$ and $F_{\sigma\rightarrow\pi}$ are
the $\sigma\rightarrow\sigma$ and $\sigma\rightarrow\pi$ scattering
amplitudes.

The dependence of polarization and intensity of the scattered
radiation on $\psi$ can be modeled with satisfactory agreement by
assuming a triple-$\vec{q}$ \emph{antiferroquadrupolar ordering} of
($\Gamma_5$) quadrupoles, and using the ATS cross section
\cite{Templeton85,Dmitrienko83,Dmitrienko84} as detailed below and
shown by the lines drawn in Fig.~\ref{fig.intensities}.

\begin{figure}
  \centerline{%
    \includegraphics[width=0.65\columnwidth]{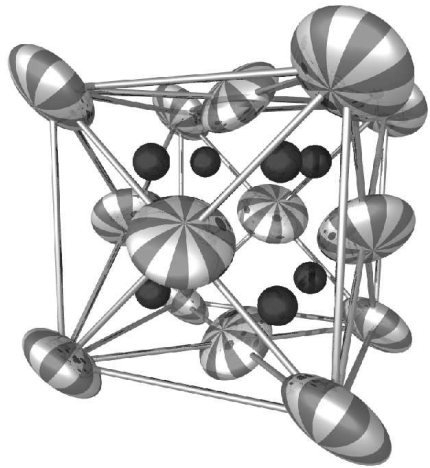}
    }
  \caption[Crystal structure of $\mathrm{NpO_2}$ in the
  antiferroquadrupolar state]{\label{fig.structure}Crystal structure
    of $\mathrm{NpO_2}$ in the antiferroquadrupolar state with space
    group $Pn\bar{3}m$. The ellipsoids represent the orientation of
    the local symmetry axis at the $\mathrm{Np}$ position, not the
    actual charge distributions. The $\mathrm{O}$ atoms are shown as
    spheres.}
\end{figure}

Above $\mathrm{T_0}$, $\mathrm{NpO_2}$ crystallizes in the Fluorite
structure with space group (SG) $Fm\bar{3}m$, $\mathrm{Np}^{4+}$ at
$4a$ and $\mathrm{O}^{2-}$ ions at $8c$ Wyckoff positions. A periodic
distribution of electric quadrupoles can be Fourier expanded, whatever
the type of order. A triple-$\vec{q}$ structure is obtained when $3$
components of the star of the propagation vector $\vec{q}$ enter in
the Fourier sum. For $\vec{q}_1=(1~0~0)$, $\vec{q}_2=(0~1~0)$ and
$\vec{q}_3=(0~0~1)$, the arrangement schematically shown in
Fig.~\ref{fig.structure} is obtained, with charge distribution
distorted along the $\langle 1~1~1 \rangle$ directions of the cubic
unit cell.  The SG symmetry is lowered from $Fm\bar{3}m$ to
$Pn\bar{3}m$, the only maximal non-isomorphic subgroup of $Fm\bar{3}m$
that is non-symmorphic and simple cubic. Within this SG, $\mathrm{Np}$
ions can be accommodated on the $4b$ positions, with reduced point
symmetry $D_{3d}$ but the same crystallographic extinction rules of
the FCC para-quadrupolar SG\@. Oxygen ions can occupy two inequivalent
positions ($2a$ and $6d$), where all coordinates are \emph{uniquely
  fixed by symmetry alone}, so that this electronic phase transition
does not allow a shift of the oxygen ions.  The symmetry of the
ordered state remains cubic and, apart from a possible change of the
lattice parameter \cite{Mannix99}, the transition will not be
accompanied by any distortion. In that case, the quadrupolar OP
\emph{cannot} be measured with neutron or \emph{conventional} X-ray
diffraction techniques.

RXS from quadrupolar order has recently been observed in several
systems \cite{Murakami98,McMorrow01}. It is well described within the
framework of ATS scattering
\cite{Templeton85,Dmitrienko83,Dmitrienko84}, which occurs when the
photon energy is tuned to an absorption edge of an atom. The
anisotropy of this atom's polarizability may lead to a finite
scattering cross section at reflections which are normally forbidden
due to glide plane or screw axis extinction rules. The scattering
amplitude arising from E1 transitions can be described by second rank
tensors, which are invariant under the point symmetry of the
scattering atom.

The scattering amplitude for the induced quadrupolar order must be
represented by a symmetric tensor $\tilde{f}(\vec{Q})$.  For
$\vec{Q}=(0~0~L)$, with $L$ odd, the scattering length is given by
$F(\vec{Q})=\vec{\epsilon}^\prime \cdot \tilde{f}(\vec{Q}) \cdot
\vec{\epsilon} = \tilde{\Phi} ( \epsilon^\prime_x \epsilon_y +
\epsilon^\prime_y \epsilon_x)$, where $\vec{\epsilon}$ and
$\vec{\epsilon}^\prime$ are the polarization vectors of the incident
and scattered beam, and $\tilde{\Phi}$ is proportional to the Fourier
component of the quadrupolar operator, $\Phi$.  With incident $\sigma$
polarization and $\vec{h}_0=(1~0~0)$ as azimuthal reference vector, we
find
\begin{eqnarray}
  F_{\sigma\rightarrow\sigma} 
  & = & 
  \tilde{\Phi} \sin(2\psi) \label{eq.sigma}\\
  F_{\sigma\rightarrow\pi}    
  & = & 
  \tilde{\Phi} \sin(\theta)\cos(2\psi), 
  \label{eq.pi}
\end{eqnarray}
where $\theta$ is the Bragg angle. The resulting scattered intensities
are $I_{\sigma\rightarrow\sigma}\propto
|F_{\sigma\rightarrow\sigma}|^{2}$ and $I_{\sigma\rightarrow\pi}
\propto |F_{\sigma\rightarrow\pi}|^{2}$.  As shown in
Fig.~\ref{fig.intensities}(a), the experimental data are well matched
by this model with one overall scale factor. This factor can be
eliminated by calculating the Stokes parameter $P_1$. The data are
shown in Fig.~\ref{fig.intensities}(b), along with model calculations.

The quantities $I_\pm$ defined in eq.~\ref{eq.one} are given by
\begin{equation}
  I_{\pm}
  \propto \left|\tilde{\Phi}\right|^2
  \left(1-\cos^2(\theta)\cos^2(2\psi)
    \pm \sin(\theta)\sin(4\psi)
  \right).
  \label{eq.plusminus}
\end{equation}
The Stokes parameter $P_2$ can be calculated, and is in good agreement
with the experiments, Fig.~\ref{fig.intensities}(c).

However, the quadrupolar order \emph{alone} is not a sufficient
ingredient, as it cannot explain the absence of a disordered magnetic
moment \cite{Erdoes80} and the breaking of invariance under time
reversal \cite{Kopmann98}. The lowest-rank multipolar OP consistent
with the experimental findings (no ordered and fluctuating dipoles,
broken time reversal symmetry) is an \emph{octupole} \cite{Santini00}.
Under octahedral symmetry, the seven octupolar operators belong to
irreducible representations $\Gamma_{2}$, $\Gamma_{4}$ or
$\Gamma_{5}$. A $\Gamma_{4}$-octupole OP must be ruled out. In fact,
this would always be accompanied by an ordering of dipoles as the
latter belong to $\Gamma_{4}$ as well. A $\Gamma_{2}$ OP is consistent
with most properties of $\mathrm{NpO_2}$ \cite{Santini00}, but it
cannot explain the RXS results, as it does not carry an electric
quadrupole moment and the $\sigma\rightarrow\sigma$ signal would
vanish\cite{Lovesey96b}. $\Gamma_{5}$ octupolar operators are given by
symmetrized combinations of $O_i = J_i ( J_j^2-J_k^2)$, where
$ijk=xyz$, $yzx$ or $zxy$. The little cogroup of the ordering
wavevector is $D_{4h}$. $\Gamma_{5}$ decomposes into the tetragonal
representations $\Gamma^{(t)}_{4}$ (1-d) and $\Gamma^{(t)}_{5}$ (2-d),
which identify two possible type-I octupolar orders, a
``longitudinal'' one and a ``transverse'' one.  For a single-$\vec{q}$
structure, both types of OPs are inconsistent with the observed
quenching of dipoles and with the present RXS results.  There is just
one single type of order which can explain all observations, and this
is a triple-$\vec{q}$ longitudinal structure, in which the
$\Gamma^{(t)}_{4}$ OPs associated with the three wavevectors of the
star of $\vec{q}$ have the same amplitude, $\rho$.  The simplest
conceivable mean-field (MF) Hamiltonian for each of the four
sublattices $s$ includes the crystal field (CF) potential
\cite{Santini00} and a self-consistent octupolar interaction term
\begin{equation}
\lambda O(\vec{n} (s)) \langle  O(\vec{n} (s))\rangle (T),
\end{equation}
where $\vec{n}(s)$ represents one of the four inequivalent %
$\langle 1~1~1 \rangle$ directions, $O(\vec{n} (s)) = \sum_{l=x,y,z}
O_l n_{l}(s)$, $\lambda$ is a MF constant, and $\langle O(\vec{n}
(s))\rangle (T)$ is the self-consistent average value of $O(\vec{n}
(s))$, which does not depend on $s$ and is nonzero for
$\mathrm{T<T_0}$. The MF potential lowers the symmetry of the
$\mathrm{Np}$ Hamiltonian from $O_h$ to $D_{3d}$ and splits the
$\Gamma_8$-quartet CF ground state into two singlets, $\Gamma_5$ and
$\Gamma_6$, and one doublet, $\Gamma_4$, of $D_{3d}$.  $\Gamma_5$ and
$\Gamma_6$ are complex-conjugate representations, which are degenerate
for a time-reversal invariant Hamiltonian, whereas their degeneracy is
lifted by the time-odd OP\@.  By choosing the value of $\lambda$ which
reproduces the observed $\mathrm{T_0}$, the level sequence at
$\mathrm{T=0}$ is found to be singlet-doublet-singlet. The ground
state has zero dipole moment and the susceptibility saturates for
$T\rightarrow 0$, as observed \cite{Erdoes80}. These properties do not
depend on the details of the CF Hamiltonian (i.e.{\ }on the precise
value of the CF parameter $x$ or whether J-mixing is taken into
account).

This octupolar OP induces the observed triple-$\vec{q}$ structure of
$\Gamma_{5}$ quadrupoles as secondary OP, with an amplitude $\propto
\rho^2$ in MF near $T_0$. The charge density is distorted along
$\vec{n} (s)$, as quantified by the quadrupolar operator along
$\vec{n} (s)$, $\Phi(\vec{n} (s)) \propto \alpha [3(\vec{n} (s)\cdot
\vec{J})^2 - J(J+1)]$ where $\alpha$ is a Stevens coefficient. $\Phi$
is negative in the ordered phase, indicating a charge distribution
``oblate" along $\vec{n} (s)$. This quadrupolar secondary OP is the
quantity observed in the RXS experiment. 


The model we propose excludes a lattice distortion or a shift of the
oxygen positions. The reduction of the local symmetry at the
$\mathrm{Np}$ site leads to an electric field gradient along the
$\langle 1~1~1 \rangle$ directions. This explains the line broadening
observed below $\mathrm{T_0}$ in M{\"o}{\ss}bauer spectroscopy
\cite{Friedt85}.  The octupolar order breaks time reversal symmetry
and thus allows the occurrence of interstitial magnetic fields as
evidenced by $\mu$SR \cite{Kopmann98}.

Finally, we remind the reader that we have \emph{inferred} the
triple-$\vec{q}$ symmetry and octupolar order from the boundary
conditions set out by previous experimental observations, such as the
absence of lattice distortions, the analogy to $\mathrm{UO_2}$, and
from the excellent agreement of our data with the model. At the E2
edge, the principal OP, {i.e.\ }the $\Gamma_{5}$ magnetic octupole, is
expected to give a resonant contribution to the scattered intensity
\cite{Lovesey96b}, comparable to that expected at the same energy from
the secondary OP, the $\Gamma_{5}$ electric quadrupoles.  However, no
signal has been detected at the E2 edge, indicating that both
contributions are below the sensitivity of the experimental device we
used. At the E1 edge the intensity is almost entirely of quadrupolar
origin, Indeed, the $\sigma\rightarrow\sigma$ intensity has no
magnetic contributions, and we can fit both the
$\sigma\rightarrow\sigma$ and the $\sigma\rightarrow\pi$ intensities
using a single scaling factor, which should not be possible with a
sizable magnetic contribution in the $\sigma\rightarrow\pi$ channel.
Such a contribution could originate from an ordered magnetic moment in
the bulk, but M{\"o}{\ss}bauer spectroscopy establish it unambiguously
to be vanishing, or from dipolar order around defects, but this would
give broad peaks which are not observed.

Whilst we have no \emph{direct evidence} for a triple-$\vec{q}$
magnetic structure, and cannot directly prove the octupolar model,
several predictions about the low temperature properties can be
investigated.  Octupolar order will split the $\Gamma_{8}$ CF ground
state into a singlet ground state and two excited levels.  Inelastic
neutron scattering \cite{Amoretti92} has shown an excitation near
$\mathrm{6.5~meV}$, but the second one is expected to lie below
$\mathrm{2~meV}$, outside the explored energy range.  Low temperature
specific heat measurements should also show the characteristic
signature of the CF spectrum.  The integrated entropy from lowest
temperatures up to the phase transition should be $R \ln(4)$, with a
significant contribution from a Schottky anomaly below $\mathrm{8~K}$.
If large single crystals would become available, octupolar order could
also be directly observed by neutron diffraction, with a form factor
peaking at a finite value of the momentum transfer and reflecting the
Fourier transform of the static magnetization density below $T_0$.

\acknowledgments

We thank the ESRF and the beamline staff on ID20 for help with the
experiments. Discussions with G.  Amoretti and R. Walstedt are
gratefully acknowledged.


\end{document}